\documentclass[aps,floatfix, showpacs,twocolumn, amsmath, amssymb,  nofootinbib]{revtex4}
\usepackage{dsfont}
\usepackage{epsfig}
\usepackage{graphicx}
\usepackage{slashed}
\usepackage{color}
\usepackage{subfigure} 
\usepackage{lineno,hyperref}
\newcommand{\bda}{\begin{\displaymath}\begin{array}{rl}}
\newcommand{\eda}{\end{array}\end{displaymath}}
\newcommand{\be}{\begin{equation}}
\newcommand{\ee}{\end{equation}}
\newcommand{\beq}{\begin{equation}}
\newcommand{\eeq}{\end{equation}}
\newcommand{\bdm}{\begin{displaymath}}
\newcommand{\edm}{\end{displaymath}}
\newcommand{\bea}{\begin{eqnarray}}
\newcommand{\eea}{\end{eqnarray}}
\newcommand{\gev}{\, \mbox{\rm GeV}}
\newcommand{\mev}{\, \mbox{\rm MeV}}

\newcommand{\fm}{\, \mbox{\rm fm}}

\newcommand{\rs}{\langle r^2\rangle\rule[-0.2em]{0em}{0em}_s^\pi}

\begin{document}
~\vspace{1cm}

\title{ Model-independent constraint on the pion scalar form factor and light quark masses}

\author{Irinel Caprini}
\affiliation{Horia Hulubei National Institute for Physics and Nuclear Engineering,
P.O.B. MG-6, 077125 Bucharest-Magurele, Romania}

\begin{abstract}
We investigate the pion scalar form factor in the Meiman-Okubo framework,  implementing the phase  below the inelastic $K\bar K$ threshold, where it is known  from the $\pi\pi$ scalar isoscalar  phase shift $\delta_0^0$ by Watson theorem. State-of-the-art knowledge of the perturbative QCD expansion of the scalar correlator and the phase shift $\delta_0^0$  is used as input. No assumptions about the phase above the inelastic threshold or the possible zeros of the form factor in the complex plane are necessary.  We obtain a  model-independent constraint relating the sum of the light quark masses to the slope and the curvature of the pion scalar form factor at the origin.  The recent lattice results for the light quark masses and the pion scalar radius are found to satisfy this  constraint.  We obtain also a strong correlation between the pion scalar radius and the curvature of the form factor, with rather high values  predicted for the curvature. 
\end{abstract}


\pacs{11.55.Fv, 11.30.Rd, 13.75.Lb}
\maketitle

\section{Introduction}\label{sec:intro}
The pion scalar form factor  $\Gamma_\pi(t)$ is defined by the matrix element 
\be\label{eq:def}
 \langle \pi^a(p)\pi^b(p') \,{\rm out}|S(0)  |0\rangle = \delta^{ab} \Gamma_\pi(t)\,,\quad t=(p+p')^2\,,
\ee
of the scalar operator
\be\label{eq:S}
S(x)=\hat m [\bar{u}(x) u(x)+ \bar{d}(x) d(x)]\,,\quad \hat m \equiv \frac{1}{2}(m_u+m_d)\,,
\ee
where $u, d$ are quark fields and $m_u, m_d$ the quark current masses.
 
 Since the Higgs boson is not light, the pion scalar form factor is not accessible to experiment. However, it
is important for theory, reflecting crucial aspects of QCD at low energy. Its Taylor expansion at $t=0$:
\be\label{eq:Taylor}
\Gamma_\pi(t)=\Gamma_\pi(0)\left[1+\frac{1}{6} \rs t +c_s^\pi t^2+\dots\right]\,,
\ee 
convergent in a  disk limited by the nearest branch point of  $\Gamma_\pi(t)$  at $t=4 m_\pi^2$, has been investigated in chiral perturbation theory  ($\chi$PT), where the pion scalar form factor has been calculated up to two loops \cite{GaLe, GaMe, Bijnens Colangelo Talavera 1998}. 

 The value of the form factor at zero momentum transfer, $\Gamma_\pi(0)$, 
 referred to as the pion $\sigma-$term, describes the dependence of the pion mass on the quark masses and was evaluated using
 Gell-Mann-Oakes-Renner relation \cite{GMOR}.  The  value quoted in \cite{GaMe} is
\be\label{eq:FH}
\Gamma_\pi(0)=(0.99\pm 0.02)\, m_\pi^2 +O(m_\pi^6).
\ee
The uncertainty in this relation   might be somewhat underestimated, since Ref. \cite{GMOR} 
included only parts of the higher-order corrections. We shall discuss in the last section
the impact of a larger uncertainty on the results derived in this paper. 

The quadratic scalar radius $\rs=6 \Gamma_\pi'(0)/\Gamma_\pi(0)$ 
 is   connected to another important quantity of $\chi$PT, the effective chiral constant
  $\bar{l}_4$ that determines the first
  nonleading contribution in the chiral expansion of the pion decay constant $f_\pi$. It also contributes to the S-wave  $\pi\pi$  scattering lengths $a_0^0$ and $a_0^2$ \cite{CGL}.

 From general principles it is known  that  $\Gamma_\pi(t)$ is an  analytic function of hermitian (real) type (i.e., it satisfies the Schwarz reflection relation $\Gamma_\pi(t)^*=\Gamma_\pi(t^*)$) in the $t$ complex plane with a cut determined by unitarity for  $t\ge 4 m_\pi^2$.   Watson final-state theorem states that below the  first inelastic threshold, which in practice is due to the $K\bar K$ channel, the phase of the form factor is equal to   the phase-shift $\delta_0^0(t)$ of the $I=L=0$ partial-wave amplitude of $\pi\pi$ elastic scattering: 
\be\label{eq:Watson}
\arg[\Gamma_\pi(t+i \epsilon)]=\delta_0^0(t)\,, \quad 4 m_\pi^2 \le  t \le 4m_K^2\,.
\ee

There have been some discussions in the literature about the value of $\rs$ obtained in the frame of dispersion theory. While the treatments \cite{DGL, CGL, Moussallam 1999} based on  Mushkhelishvili-Omn\`es equations give
$\rs$ in the range $(0.57$ - $0.65) \fm^2$,  the calculations \cite{Ynd1, Ynd2} based on single-channel Omn\`es formalism led to a higher prediction,  $\rs=(0.75 \pm 0.07) \fm^2$. This discrepancy was discussed in 
\cite{ACCGL}, where it was shown that the single-channel treatment  can be made consistent with  the multi-channel one if one takes into account the fact that Watson theorem is valid modulo $\pm \pi$ (an interpretation in terms of a possible zero of the form factor was given in \cite{Oller}). In this context, any alternative investigation 
of the scalar form factor based on analyticity, which might improve the knowledge of  the scalar radius, is of great interest.

In a recent paper \cite{PRL:2017}, a precise determination of the charge radius of the pion was obtained in the frame of a mixed dispersion formalism, using as input the phase of the electromagnetic form factor in the elastic region and the modulus above the first inelastic threshold. Unfortunately, in the case of the scalar form factor (\ref{eq:def}) no data on the modulus  above the $K\bar K$ threshold are available. One can obtain  however an integral constraint on the modulus squared of $\Gamma_\pi(t)$ along the cut using a formalism proposed a long time ago by Meiman \cite{Meiman:1963} and Okubo \cite{Okubo:1971jf}, which exploits the dispersion relations for suitable QCD correlators, combined with unitarity and the  positivity of the spectral functions. 
This formalism has been applied in the context of QCD for the first time in \cite{Bourrely:1980gp}, and afterwards in  many papers \cite{deRafael:1992tu}-\cite{CaGrLe}, being in particular a valuable tool for obtaining model-independent constraints on weak semileptonic form factors.

The Meiman-Okubo formalism has been applied also  to the light-quark scalar correlator in \cite{LeRa, LL:proc}, where it was used as a mean to derive a lower bound on the sum of the light quark masses. In the present paper we revisit the analysis reported in \cite{LeRa, LL:proc}   bringing  several improvements and updates. Thus, while in these works Watson theorem (\ref{eq:Watson}) was implemented only below 0.5 GeV, now it can be imposed up to the first relevant inelastic threshold, set by the $K\bar K$ channel, taking advantage of the recent progress in the determination of the pion-pion phase shifts. We include also higher terms in the expansion (\ref{eq:Taylor}), which will lead to a more general constraint on the sum of the light quark masses, the derivatives of the scalar form factor at the origin and the  phase shift $\delta_0^0$.  Finally, we use the most recent calculation of the scalar correlator in perturbative QCD,  available now to $O(\alpha_s^4)$ \cite{BaChKh}. The motivation of revisiting this analysis  is the fact that  precise results for both  the light-quark masses and the pion scalar radius are now available from lattice calculations (for a recent review and earlier references see \cite{FLAG}). An updated, more precise independent constraint  on these quantities is therefore of interest. 

In the next section we describe the mathematical formalism, in Sec. \ref{sec:input} we discuss the input used in the calculations and in Sec. \ref{sec:res} we present our results. Sec. \ref{sec:conc} contains a discussion of the results in comparison with previous determinations and our conclusions.
 
\section{Derivation of the bounds}\label{sec:math}
We consider the scalar correlator \cite{LeRa}
\be\label{eq:Psi}
\Psi(q^2)=i \int dx e^{i q\cdot x} \langle 0|T(S(x)S^\dagger(0))|0\rangle\,,
\ee
written in terms of the operator $S(x)$ defined in (\ref{eq:S}).

The function $\Psi(q^2)$ satisfies a dispersion relation which  requires two subtractions. Therefore, as in \cite{LeRa} we shall consider the second derivative of $\Psi$, which is expressed on the  Euclidian axis,  i.e., for $Q^2=-q^2>0$, as
\be\label{eq:DR}
\Psi''(Q^2)=\frac{1}{\pi}\int_0^\infty \frac{2}{(t+Q^2)^3}\, \mbox{\rm Im} \Psi(t)\,dt\,.
\ee
At large $Q^2$, $\Psi''(Q^2)$ is given by the QCD perturbative expansion in powers of the  renormalized strong coupling $\alpha_s$  with negligible power corrections \cite{BaChKh}:
\be\label{eq:OPE}
\hspace{-0.2cm}\Psi''(Q^2)=\frac{3}{16 \pi^2} \frac{(m_u+m_d)^2}{Q^2}\, \left[1+  \sum_{n\ge 1} \bar{d}_{0, n} \left(\frac{\alpha_s}{\pi}\right)^n  \right]\!,
\ee
 where the quark masses and the strong coupling are evaluated at a fixed scale $\mu^2$.  

On the other hand, on  the timelike axis the spectral function  $\mbox{\rm Im} \Psi(t)$ can be expressed, using unitarity, in terms of the contributions of  hadronic states. Keeping the  lowest two-pion contributions and using the positivity of the spectral function one obtains \cite{LeRa}
\be\label{eq:unit}
\mbox{\rm Im} \Psi(t)\ge\frac{3}{16\pi}\sqrt{1-\frac{4m_{\pi}^2}
{t}}\,\vert \Gamma_\pi(t)\vert^{2}\theta(t-4m_{\pi}^2),
\ee 
where  $\Gamma_\pi(t)$ is the pion scalar form factor defined in (\ref{eq:def}).

From (\ref{eq:DR}) and (\ref{eq:unit})  one obtains the inequality
\be\label{eq:ineq}
\Psi''(Q^2)\ge \frac{3}{8\pi^2}\int_{t_p}^\infty \frac{1}{(t+Q^2)^3}\sqrt{1-t_p/t}\,\vert \Gamma_\pi(t)\vert^{2}\,dt,
\ee
where we denoted $t_p=4 m_\pi^2$.  

By applying standard techniques of complex analysis, the right hand side of (\ref{eq:ineq}) can be further bounded from below by definite expressions involving values of the form factor at points inside the analyticity domain or the coefficients of the Taylor expansion  (\ref{eq:Taylor}) at $t=0$. 

In order to derive the optimal lower bound on the rhs of (\ref{eq:ineq}) with the constraints (\ref{eq:Taylor}) and  (\ref{eq:Watson}), we first perform the conformal mapping 
\be\label{eq:z}
z\equiv \tilde z(t)=\frac{1-\sqrt{1-t/t_p}}{1+\sqrt{1-t/t_p}},
\ee
which maps the complex $t$ plane cut for $t\ge t_p$ onto the unit disk $|z|<1$ such that $\tilde z(0)=0$, $\tilde z(t_p)=1$ and the upper (lower) edge of cut becomes the unit  semicircle $\zeta=e^{i \theta}$ with $\theta>0$ ( $\theta<0$).  Then  the inequality (\ref{eq:ineq}) can be written in the equivalent form
\be\label{eq:ineqz}
\Psi''(Q^2)\ge \frac{1}{2 \pi} \int_{-\pi}^{\pi} |\phi(\zeta)\, \Gamma_\pi(\tilde t(\zeta))|^2 d\theta\,,
\ee
where   $\tilde t(z)=4 t_p z/(1+z)^2$ is the inverse of (\ref{eq:z}) and $\phi(z)$ is an analytic function in $|z|<1$ defined as
\be\label{eq:phi}
\phi(z)=\sqrt{\frac{3}{4 \pi}}\,\frac{1}{t_p}\,\frac{(1-z)(1+z)^{3/2}}{(1-z+\beta_Q (1+z))^3}\,,
\ee
with $\beta_Q=\sqrt{1+Q^2/t_p}$. By construction, $\phi(z)$ in an ``outer function'' \cite{Duren},  i.e., it has no zeros in $|z|<1$, and its modulus on the boundary $|z|=1$ is given by 
\be
|\phi(\zeta)|^2=\frac{3}{8\pi}\frac{\sqrt{1-t_p/\tilde t(\zeta})}{(\tilde t(\zeta)+Q^2)^3} \left|\frac {d\tilde t(\zeta)}{d\zeta}\right|\,.
\ee
We emphasize that the absence of zeros in $\phi(z)$ guarantees that the bounds derived below are optimal. 

We further define a new function $g(z)$, analytic in $|z|<1$, by
\be\label{eq:g}
g(z)=\phi(z) \,\Gamma_\pi(\tilde t(z))\,.
\ee
Then  (\ref{eq:ineqz})  implies the following inequality, valid for any $K\ge 1$,
\be\label{eq:domain0}
\Psi''(Q^2)\ge \sum_{k= 0}^{K-1}\, g_k^2\, ,
\ee
where the real coefficients $g_k$ are defined by the Taylor expansion
\be\label{eq:gz}
g(z)=\sum_{k= 0}^\infty g_k\, z^k\,.
\ee
From (\ref{eq:g}) and (\ref{eq:gz}) it follows that each $g_k$ is a linear combination of the derivatives of $\Gamma_\pi(t)$ at $t=0$  up to the order $k$.

We now improve (\ref{eq:domain0}) by taking into account the additional constraint (\ref{eq:Watson}). We apply  a technique of functional optimization based on functional Lagrange multipliers,  proposed for the first time in \cite{Micu:1973vp}, generalized and applied further in many papers \cite{IC, Bourrely:2005hp, Abbas:2010jc, CaGrLe}. We write below the solution of this optimization problem in our case (for a proof see \cite{IC, Abbas:2010jc}). Let
\beq
\zeta_{\rm in} \equiv \tilde z(4 m_K^2)= e^{ i\theta_{\rm in}}
\eeq
be the image on the unit circle in the $z$ plane of the point
$4 m_K^2+i\epsilon$ situated on the upper edge of the cut [the point
$4 m_K^2-i\epsilon$ being mapped onto $\exp(-i\theta_{\rm in}$)].  Then one obtains
 the stronger condition \cite{IC, Abbas:2010jc}
\beq\label{eq:domain}
\hspace{-0.1cm}\Psi''(Q^2)\ge \sum_{k = 0}^{K-1} g_k^2+ \sum_{k = 0}^{K-1} \frac{g_k}{\pi}\!\!
\int\limits_{-\theta_{\rm in}}^{\theta_{\rm in}} \!\!
 d \theta \, \lambda(\theta) 
\sin\left[k \theta - \Phi(\theta) \right],
\eeq
where the function $\lambda(\theta)$  is the solution of the integral equation
\beq\label{eq:eq1}
 \hspace{-0.2cm}\sum_{k = 0}^{K-1} g_k \sin[k \theta - \Phi(\theta)] =\lambda(\theta)
 - \frac{1}{2\pi} \!\! \int\limits_{-\theta_{\rm in}}^{\theta_{\rm in}}\!
 {\rm d} \theta' \lambda(\theta') \, {\cal K}_{\Phi}(\theta, \theta'),
\eeq
valid for $\theta \in (-\theta_{\rm in}, \theta_{\rm in}) $, where the
kernel is defined as
\beq\label{eq:calK}
{\cal K}_{\Phi}(\theta, \theta') = \frac{\sin[(K-1/2) (\theta -
\theta') - \Phi(\theta) +
\Phi(\theta^\prime)]}{\sin[(\theta-\theta^\prime)/2]},
\eeq
in terms of the function 
\beq\label{eq:Phi}
 \Phi(\theta)= \delta_0^0(\tilde t(e^{i\theta})) +\arg[\phi(e^{i\theta})].
\eeq

We mention that integral equations of the form (\ref{eq:eq1})  are often encountered in solving functional optimization problems with constraints on the boundary implemented by Lagrange multipliers. The function $\lambda$ (which is actually a generalized Lagrange multiplier) is a smooth, odd function of $\theta$ defined on the range $(-\theta_{\rm in}, \theta_{\rm in})$, with values depending on the free parameters $(r_\pi, c_\pi)$  used in the optimization procedure.  If the function $\Phi(\theta)$ defined in (\ref{eq:Phi}) is sufficiently smooth, the integral equation (\ref{eq:eq1}) is of Fredholm type and is solved numerically by approximating it by a linear system of equations obtained by discretizing the integral. In the calculations performed in this work, the results proved to be very stable when the number of integration points was increased up to several hundreds.

The inequality (\ref{eq:domain}), with $\Psi''(Q^2)$, given by (\ref{eq:OPE}), provides a model-independent relation between the sum of the light quark masses, the coefficients of the Taylor expansion of the scalar form factor, and its phase  on the elastic part of the unitarity cut. In Sec. \ref{sec:res} we shall explore numerically the consequences of this result. Before this we  
review in the next section the quantities used as input in the calculations.
\section{Input}\label{sec:input} 
 The perturbation expansion (\ref{eq:OPE}) of the scalar correlator  is known at present  in the $\overline{\rm MS}$ scheme to $O(\alpha_s^4)$ \cite{BaChKh}. We used the coefficients $\bar{d}_{0,j}$ written in \cite{BaChKh,ChKh} for $1\le j\le 4$ as polynomials of degree $j$ of the quantity $L_{Q}= \log Q^2/\mu^2$.

\begin{figure}[ht!]
\begin{center}
\includegraphics[scale=0.35]{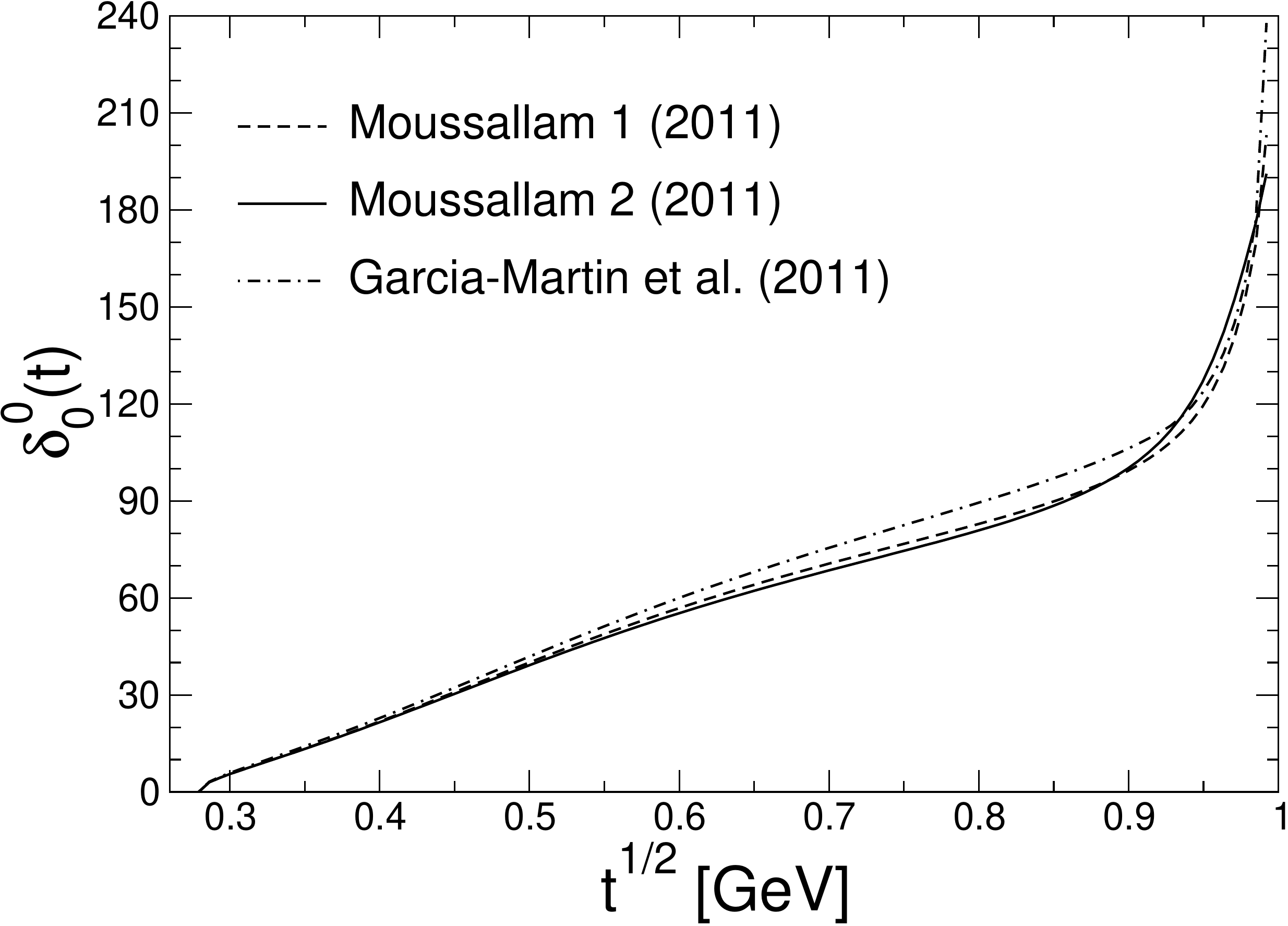}
\end{center}
\caption{Phase shift $\delta_0^0(t)$ below the $K\bar K$ threshold calculated from Roy equations in Refs. \cite{Moussallam:2011} and \cite{GarciaMartin:2011cn}. \label{fig:1}}
\end{figure}

We have to specify the spacelike value $Q^2>0$ to be taken in the dispersion relation (\ref{eq:DR}). The obvious requirement is  that  the perturbative QCD expansion  (\ref{eq:OPE}) must be meaningful. As discussed in \cite{LeRa} and in other applications of this formalism \cite{IC, Hill:2006bq, Abbas:2010jc,  Abbas:2010ns}, the choice $Q=2 \gev$ is very reasonable for correlators involving light-quark operators, so we make this choice here. For illustration, as in \cite{LeRa}, we take also $Q=1.5 \gev$, which is still large enough compared to $\Lambda_{\rm QCD}\sim 0.300 \gev$. Other choices of $Q$ will be discussed in the last section. For the renormalization scale we made the choices $\mu=2 \gev$ and  $\mu=Q$.

 We  obtained the strong coupling $\alpha_s(\mu^2)$ by using as input $\alpha_s(m_\tau)=0.330\pm 0.010$ \cite{PDG} and evolving it to the scale $\mu$ by the renormalization group equation with $\beta$ function to the same accuracy as the correlator \cite{LaRi, Czakon}.  The quark masses at $\mu=Q$ have been calculated starting from $\mu=2 \gev$ and evolving them with the running to four loop from \cite{ChRe}. The apparent convergence of the expansion of $\Psi''(Q^2)$ is quite good: for $Q=2 \gev$ the contribution of the higher corrections in  (\ref{eq:OPE}) is of 10\% for $\alpha_s^2$ terms,  5\% for $\alpha_s^3$ and  3\% for $\alpha_s^4$. For $Q=1.5 \gev$ the corrections are larger by a factor of about two\footnote{As for other QCD correlators, the expansion (\ref{eq:OPE}) is most probably a divergent series. In the spirit of asymptotic series, we attached to it an uncertainty equal to the last term kept in the expansion.}.

Considerable progress in the calculation of the $\pi\pi$ phase shifts using experimental data and Roy equations \cite{Roy:1971tc} has been achieved recently. In particular, several determinations of $\delta_0^0(t)$ have been performed  \cite{ACGL, Caprini:2011ky, Moussallam:2011, GarciaMartin:2011cn}. We used in the calculations the two solutions given in Eq. (11) and the Appendix  of \cite{Moussallam:2011} and the most precise  solution, with CFD parameters, specified in Eq. (A.3) and  Table V of \cite{GarciaMartin:2011cn}. As can be seen from  Fig. \ref{fig:1}, where the central values of these phases are shown for $\sqrt{t}<2 m_K$, there are some differences between them, the prediction from  \cite{GarciaMartin:2011cn} exhibiting slightly larger values around 0.8 GeV and  a more pronounced increase near the $K\bar K$ threshold.   We mention that the phase-shift $\delta_0^0$ determined in  \cite{Caprini:2011ky},  with the boundary values given in Eq. (72) of that work, is quite close to the second solution of \cite{Moussallam:2011}, both exhibiting in particular a more moderate increase near the opening of the $K\bar K$ channel.

\begin{figure}[htb]
\begin{center}
\includegraphics[scale=0.28]{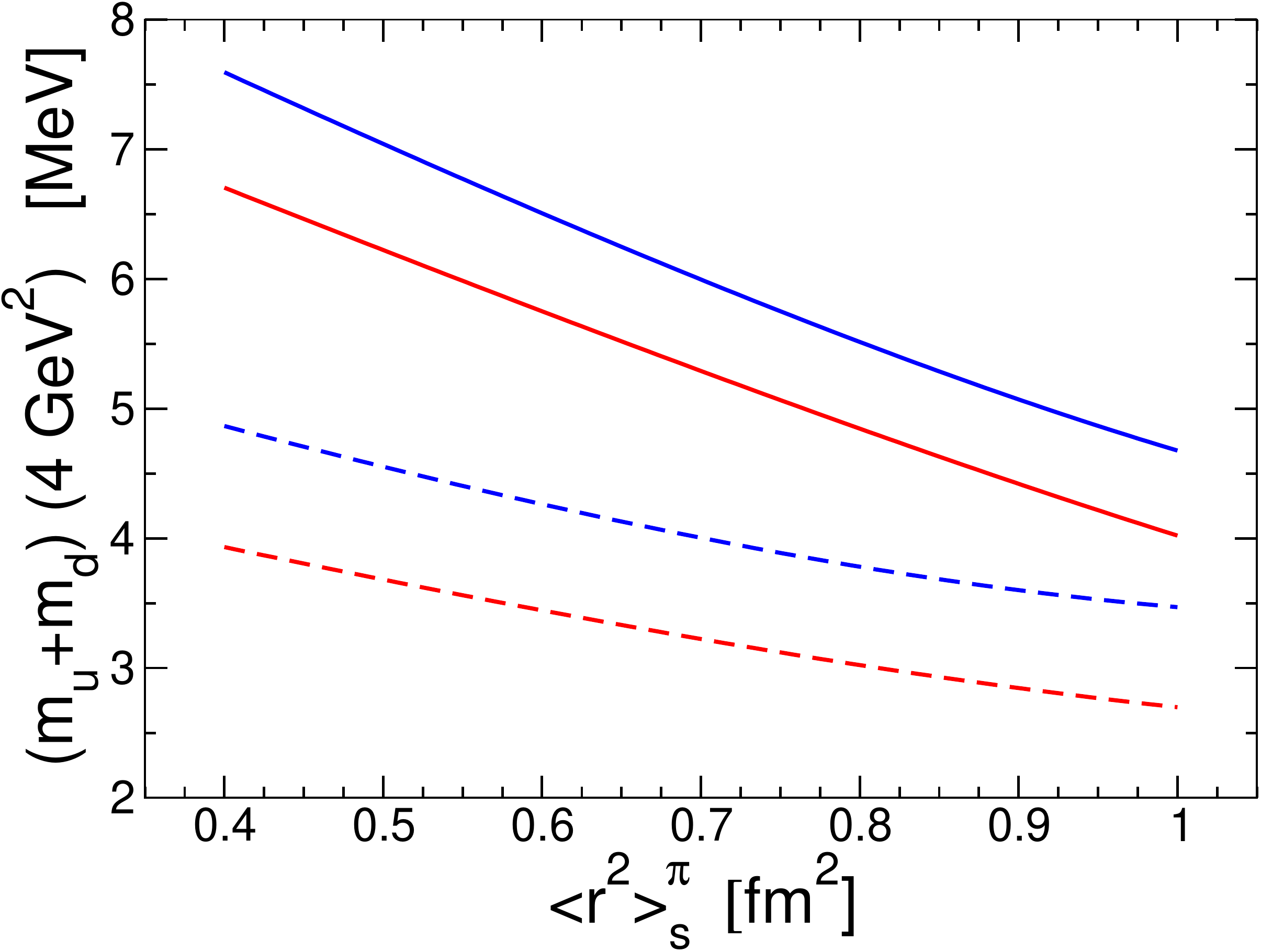}  
\caption{Boundaries of the domains  (\ref{eq:domain0}) and (\ref{eq:domain}) leading to lower bounds on $m_u+m_d$ for fixed $\langle r^2\rangle_s^\pi$ or to lower bounds on  $\langle r^2\rangle_s^\pi$ for fixed  $m_u+m_d$. Solid red (blue) lines: curves obtained with input phase equal to Solution 1 from \cite{Moussallam:2011}, for $Q=2 \gev$ ($1.5 \gev$).  Dashed red (blue) lines: curves obtained without phase input, for $Q=2 \gev$ ($1.5 \gev$). \label{fig:2}}
\end{center}
\end{figure}
\section{Results}\label{sec:res}

By first setting $K=2$ in the inequalities (\ref{eq:domain0}) and (\ref{eq:domain}) and fixing the normalization  $\Gamma_\pi(0)$,  we obtain from these inequalities a lower bound on the sum $m_u+m_d$  in terms of the scalar radius.  
Alternatively,  the inequalities impose constraints on $\rs$ for input values of quark masses. Since the right hand sides of  (\ref{eq:domain0}) and (\ref{eq:domain}) are quadratic convex functions of $\rs$,  they lead to allowed ranges for this quantity, situated between a lower and an upper bound. 

In Fig. \ref{fig:2} we show the parts of the boundaries of the domains (\ref{eq:domain0}) and (\ref{eq:domain}) leading to lower bounds on $m_u+m_d$ for fixed $\rs$, or lower bounds on $\rs$ for fixed  $m_u+m_d$, for a range of $\rs$ of physical interest. The renormalization scale has been fixed at $\mu=2 \gev$.  The allowed values of the variables are situated above the curves.  The parts of the  boundaries giving upper bounds on $\rs$ involve values of this quantity too large to be of interest and are not shown in this figure (they will appear however in Figs. \ref{fig:3}, \ref{fig:4} shown below). 

The comparison of the dashed and solid curves, obtained without and with phase input, respectively, prove the significant improvement brought by the implementation of Watson theorem. Fig \ref{fig:2} shows also that the bounds  obtained with  $Q=1.5 \gev$ (blue curves) are more stringent than those obtained with  $Q=2 \gev$ (red curves).

 The results shown in Fig.  \ref{fig:2}  have been obtained  with the central value $\Gamma_\pi(0)=0.99 \,m_\pi^2$   and  the first solution  for the phase-shift $\delta_0^0$ given in \cite{Moussallam:2011}. The phase shift from \cite{GarciaMartin:2011cn} leads to very close and slightly weaker results, while the second solution given in  \cite{Moussallam:2011} leads to somewhat better (higher) bounds. It appears that a phase-shift $\delta_0^0$ with a moderate increase near the $K\bar K$ threshold, such as exhibited by solution 2 of  \cite{Moussallam:2011} and the phase-shift calculated in \cite{Caprini:2011ky}, leads to stronger lower bounds. 
These constraints can be used for testing the consistency of specific values for the sum of light quark masses and the pion scalar radius.

 The study of the light  quarks masses has a long history in the frame of low-energy effective theory of QCD \cite{GL_QM, HL:1996}. The  quantity difficult to estimate is the difference $m_u-m_d$, or the ratio $m_u/m_d$, for which recently an accurate value was obtained through the dispersive analysis \cite{CLLP:2017}  of the isospin-breaking decay $\eta\to 3\pi$. 
We concentrate in our discussion on
the recent lattice calculations of the quark masses and $\rs$, summarized in the review \cite{FLAG}. From Table 7 of \cite{FLAG} one can obtain 13 lattice predictions for the sum   $m_u+m_d$ at the scale $\mu=2\gev$, ranging from $(5.60 \pm 0.55) \mev$ \cite{Aubin:2004fs, Aubin:2004ck} to $(8.51 \pm 0.51) \mev$ \cite{Blum:2007cy}, while four values of $\rs$ are given in Table 22 of the same review.   

Using as input the lowest of these values,  $\rs=(0.481 \pm 0.062) \fm^2$, obtained by the HPQCD Collaboration \cite{Koponen:2015tkr},  and varying the normalization  in the range (\ref{eq:FH})  we obtained with the phase 2 from \cite{Moussallam:2011} a lower bound on $m_u+m_d$ in the range $(5.99-6.86) \mev$  for $Q=2 \gev$, and in the range 
$(6.76-7.74) \mev$  for $Q=1.5 \gev$.
The range  obtained with 
$Q=1.5 \gev$ is in slight tension with the lowest value of $m_u+m_d$ in Table 7 of \cite{FLAG}, quoted above,   obtained by MILC Collaboration \cite{Aubin:2004fs, Aubin:2004ck}.   

For the other values of $\rs$ listed in Table  22 of \cite{FLAG}, which are larger than the HPQCD prediction, one obtains from  Fig. \ref{fig:2} smaller lower bounds on $m_u+m_d$, which are in no conflict with the values given in Table 7 of \cite{FLAG}. Adopting the central value $\rs=0.61 \fm^2$ of the lattice calculations in \cite{Gulpers:2015bba, Gulpers:2013uca, JLQCD:2009qn} and using as input for the phase the solution 2 from  \cite{Moussallam:2011}, the most conservative lower bound, obtained with   $Q=2 \gev$ and $\Gamma_\pi(0)=(0.99 \pm 0.02) \,m_\pi^2$, is
\be\label{eq:mfinal}
m_u+m_d\ge 5.68 \mev.
\ee
 This bound is consistent with all the lattice values listed in Table 7 of \cite{FLAG} and the average
$m_u+m_d=(6.85 \pm 0.25)\,\mev$ quoted in the Review of Particle Physics \cite{PDG}.

We can generalize  the constraints by setting in  (\ref{eq:domain0}) and (\ref{eq:domain}) the parameter $K=3$,  i.e., including also a curvature term in the expansion (\ref{eq:Taylor}).
Then, for a fixed value of the sum of quark masses, the inequalities (\ref{eq:domain0}) and (\ref{eq:domain}) describe  ellipses in the plane $\rs - c_s^\pi$. We show for illustration in Fig. \ref{fig:3} the ellipses obtained  using as input for the phase the solution 1 from \cite{Moussallam:2011} and the average value  $m_u+m_d=6.85 \mev $, quoted in \cite{PDG}.   The CFD parametrization \cite{GarciaMartin:2011cn} of the phase-shift  leads to slightly larger domains, while for the solution 2 of $\delta_0^0$ given in \cite{Moussallam:2011} the allowed domains are slightly smaller.

\begin{figure*}[ht!]
\begin{center} 
\hbox{
\includegraphics[scale=0.35]{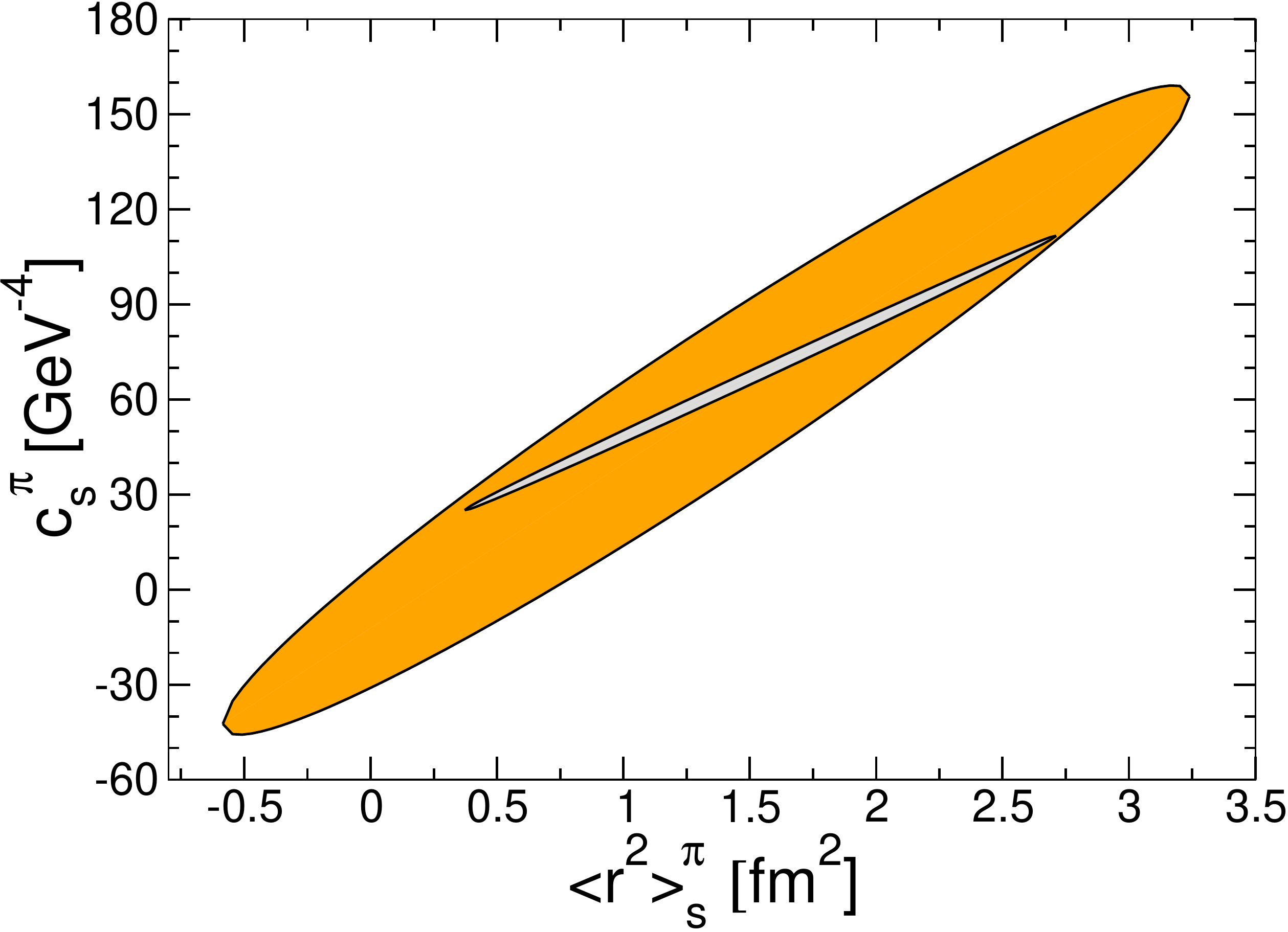}
\includegraphics[scale=0.35]{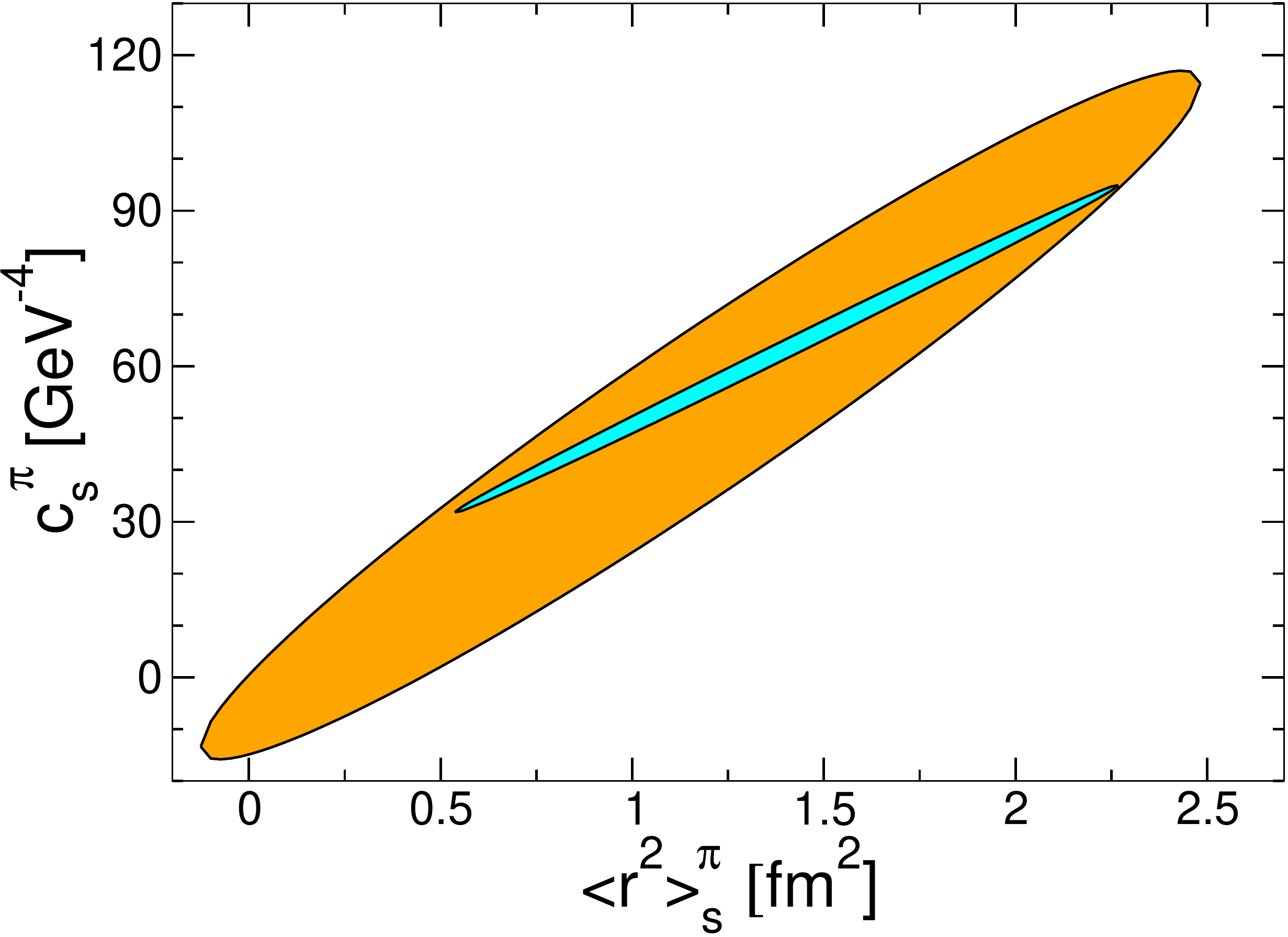}}
\end{center}
\caption{Allowed domain in the $\langle r^2\rangle\hspace{-0.01cm}_s^\pi - c_s^\pi$ plane for $m_u+m_d=6.85 \mev$ \cite{PDG}.
 Left panel:  $Q=2 \gev$. Right panel: $Q=1.5 \gev$ (note the different scales). Large ellipses: allowed domains obtained without phase input. Inner ellipses:  allowed domains with phase input from \cite{Moussallam:2011}. \label{fig:3}}
\end{figure*}

The comparison of the large ellipses with the small ones shows again the considerable effect of the implementation of Watson theorem. As before, the choice $Q=1.5 \gev$ leads to stronger constraints. In Fig. \ref{fig:4} we show for comparison the small ellipses, obtained with implementation of Watson theorem,  for  $Q=1.5 \gev$ and  $Q=2 \gev$.

From these ellipses one can read the values of the upper and lower bounds on $\rs$ obtained by using as input the sum of the quark masses. The upper bounds turn out to be very weak, so they are not useful for improving the accuracy of the dispersive predictions. On the other hand, as it was clear already from the previous discussion, the lower bounds are nontrivial. For instance, by varying the normalization (\ref{eq:FH}) we obtain the conservative lower bounds $\rs \ge 0.37 \fm^2$ for $Q=2 \gev$ and $\rs \ge 0.53 \fm^2$ for $Q=1.5 \gev$. 

Another interesting property illustrated in Figs. \ref{fig:3}, \ref{fig:4} is the strong correlation between the radius and the curvature at $t=0$. 
In particular, for $\rs=0.61 \fm^2$, we obtained for the curvature the allowed range 
\be\label{eq:c}
c_s^\pi\in (32.4-35.4) \gev^{-4},
\ee
where we included the small variations due to the uncertainties of the phase $\delta_0^0$  below the $K\bar K$ threshold, the normalization (\ref{eq:FH}), the strong coupling $\alpha_s$ and the truncation of the QCD expansion (\ref{eq:OPE}). 

The allowed values given in (\ref{eq:c}) are higher than the prediction $c_s^\pi \approx 10 \gev^{-4}$  of the coupled-channel dispersive formalism \cite{GaMe, Moussallam 1999}. A large value of the curvature  would
lead to rather large values of the relevant low energy constants (LEC) of $\chi$PT both for the two and
three-flavour case. The study of these implications is beyond the scope of the present paper. It is important however to identify possible sources of  systematic uncertainties  that can affect both  the dispersive approach and the bounds derived in this paper. We shall briefly discuss this problem  in the next section.

\begin{figure}[htb]
\begin{center}
\includegraphics[scale=0.35]{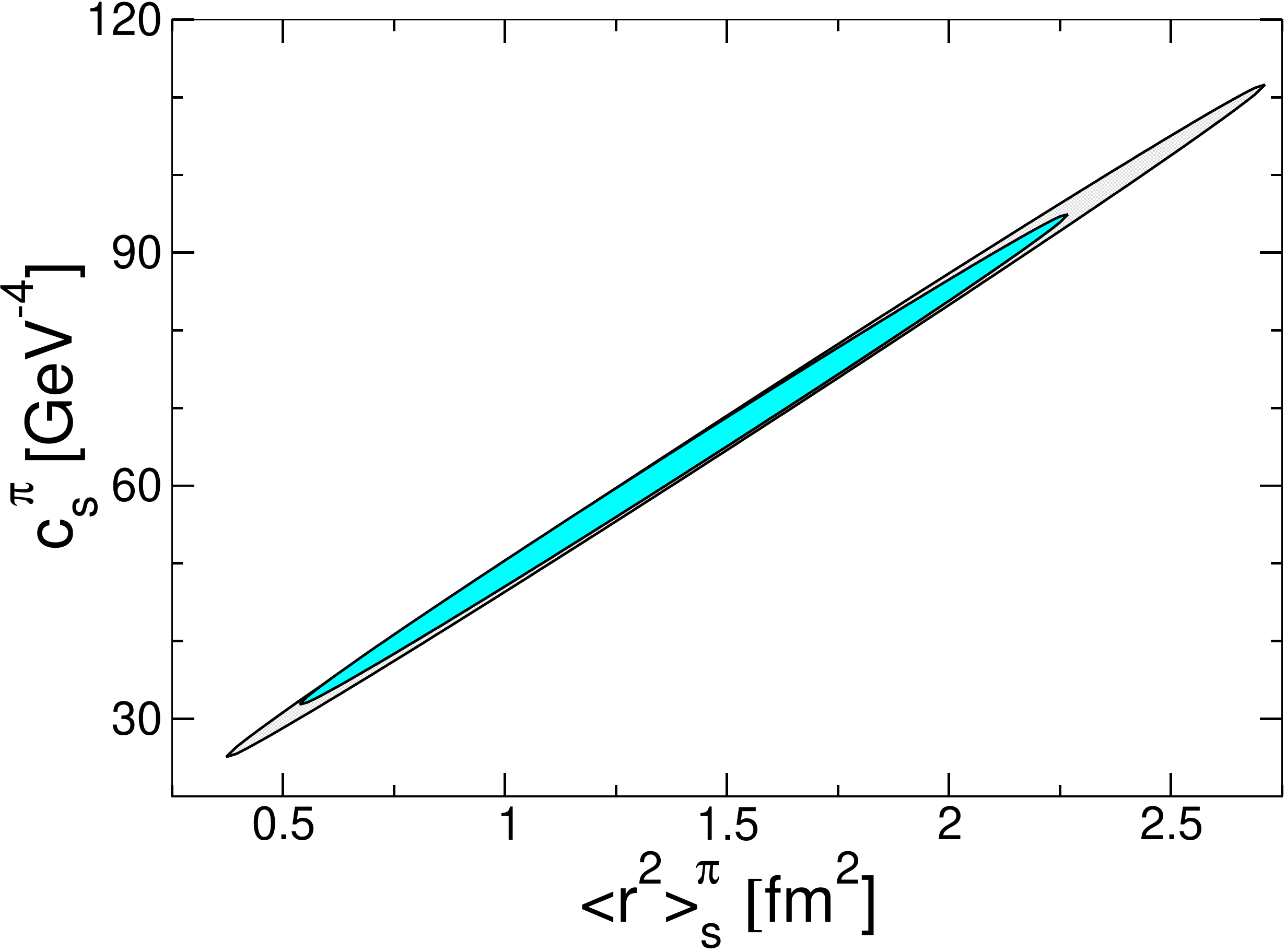}
\end{center}
\caption{ Inner ellipses from Fig. \ref{fig:3}. Gray: allowed domain for $Q=2 \gev$. Cyan: allowed domain for $Q=1.5 \gev$.\label{fig:4}}
\end{figure}

\section{Discussion and conclusions}\label{sec:conc}
In this work, we revisited the application of the Meiman-Okubo formalism to the light-quark scalar correlator, improving the previous analysis \cite{LeRa}  by the implementation of Watson theorem (\ref{eq:Watson}) up to the $K\bar K$ threshold and by the inclusion of higher derivatives in the Taylor expansion (\ref{eq:Taylor}) of the pion scalar form factor. Moreover, recent progress in the determination of  the scalar correlator in perturbative QCD and the phase shift  $\delta_0^0$ of pion-pion scattering has been taken into account.

Our result (\ref{eq:domain}) is a model-independent constraint relating the sum of the light quark masses $m_u+m_d$ (appearing in the expression (\ref{eq:OPE}) of the lhs) to the derivatives of the pion scalar form factor (entering the real coefficients $g_k$), and the phase shift $\delta_0^0(t)$ below the $K\bar K$ inelastic threshold (appearing in the expression (\ref{eq:Phi} of $\Phi(\theta)$). The phase of the form factor above the elastic region is not required in this approach. 
The results shown in Figs. \ref{fig:1} - \ref{fig:3} illustrate the significant improvement brought by the implementation of Watson theorem along the entire elastic region up to the $K\bar K$ threshold\footnote{If the phase is imposed only up to 0.5 GeV, as in \cite{LeRa}, the constraints are  weaker, the solid lines in Fig. \ref{fig:2} being shifted downward by about 1 MeV. Note that the bounds reported in  \cite{LeRa} are somewhat higher since they have been obtained with the one-loop expression for $\Psi''(Q^2)$. The question of higher perturbative orders  was addressed in \cite{LL:proc}, where it was found that a resummation lowers the bounds, in agreement with our results.}.

We have applied the above constraint for testing the consistency of the recent lattice results on the quark masses and the pion scalar radius $\rs$.  We found that, except a slight tension between the lowest value of $\rs$ given in Table 22 of the review \cite{FLAG} and the lowest value of $m_u+m_d$ in Table 7 of the same review, the recent lattice  determinations of the light quark masses and the pion scalar radius  satisfy the consistency test.

As illustrated in Fig. \ref{fig:4}, the sum of the light-quark masses and the phase below the inelastic threshold impose nontrivial constraints on the higher coefficients of the Taylor expansion (\ref{eq:Taylor}) of the pion scalar form factor.  The upper bounds on the scalar radius are very weak, being not useful for increasing the precision of the dispersive calculations of $\rs$.  On the other hand, the lower bounds turn out to be at the edge of the currently accepted values.

Figure \ref{fig:4} shows  also a strong  correlation between the radius $\rs$ and the curvature $c_s^\pi$. This result is not surprising: strong correlations between the higher derivatives  have been obtained also for other form factors in the frame of Meiman-Okubo formalism \cite{Caprini:1997mu, Bourrely:2005hp, CaGrLe}. In the present case, somewhat surprising is the fact that the allowed range (\ref{eq:c}) for the curvature corresponding to $\rs=0.61 \fm^2$ is  considerably higher than the predictions of the dispersive treatment \cite{GaMe, Moussallam 1999}.  It is of interest therefore to  discuss the possible systematic uncertainties that can affect the dispersive approach and the bounds derived in this paper.

The dispersive approach exploits unitarity, which relates the form factors to the meson-meson scattering amplitudes through a set of coupled homogeneous  integral equations \cite{GaMe, DGL, Moussallam 1999}. For solving these integral equations, each form factor is parametrized most generally as a polynomial $P(t)$ multiplied by an Omn\`es function $\Omega(t)$, defined in terms of the phase $\delta(t)$ on the cut by
\be\label{eq:omnes}
\Omega(t)= \exp \left(\frac {t} {\pi} \int^{\infty}_{4 m_\pi^2} dt' 
\frac{\delta(t^\prime)} {t^\prime (t^\prime -t)}\right).
\ee 
In particular, for the form factor $\Gamma_\pi(t)$ the polynomial was taken as a constant, which implies that the form factor was assumed to have no zeros\footnote{It is known that $\chi$PT predicts no zeros for $\Gamma_\pi(t)$ near the origin. However, one or more zeros located at larger distances in the complex plane, outside the range where  $\chi$PT is reliable, cannot be excluded by general arguments.}. This assumption may be too restrictive: the presence of a polynomial which multiplies the Omn\`es function can lead to different predictions outside the limited interval of the unitarity cut where the coupled-channel equations are solved.

 Another source of systematic uncertainty is the fact that the phase $\delta(t)$ is not known at higher energies, and some model-dependent assumptions about its behavior  are required in the dispersive approach. One can check that the modulus of the Omn\`es function (\ref{eq:omnes}) behaves as $t^{-\frac{\delta(\infty)}{\pi}}$  at large $t$. Therefore, if $\Omega(t)$ is multiplied by  a polynomial, the asymptotic phase $\delta(\infty)$ should be increased in order to ensure the asymptotic decrease as $1/t$ of the form factor, predicted by perturbative QCD.   Although the higher derivatives at $t=0$ are less sensitive to the phase at high energies, an anomalously large contribution of the inelastic channels and the presence of one or more zeros in the complex plane may have a significant contribution to the curvature. 

On the other hand, the bounds derived in the present paper have been obtained with no assumptions about the phase above the $K\bar K$ threshold or  the analytic expression of the form factor ({\em i.e.} the presence or absence of zeros).
Below (\ref{eq:c}) we mentioned the small uncertainties due to the various pieces of the input. We shall now consider in more detail the normalization condition (\ref{eq:FH}) and the perturbative QCD input as  possible sources of  systematic uncertainties.

As already mentioned, the error given in (\ref{eq:FH}) might be underestimated.  It is of interest to see how much would this uncertainty have to change in order to have a substantial impact on the results. We found that by increasing the error quoted in  (\ref{eq:FH}) by  a factor of 10, the lower bound (\ref{eq:c}) on the curvature decreases slightly,  becoming  $31  \gev^{-4}$. It turns out that in order to decrease the lower bound on the curvature below $20 \gev^{-4}$, a huge increase of the error by a factor of about 40  would be required. We emphasize that the value of the form factor at the origin does not appear in the dispersive 
treatment \cite{GaMe, Moussallam 1999}, which involves only the ratio $\Gamma_\pi(t)/\Gamma_\pi(0)$. 

Turning to the perturbative QCD input, we note that from the maximum modulus principle it follows that the bounds depend in a monotonous way on the value of the lhs of the inequality (\ref{eq:ineq}), in the sense that larger values of $\Psi''(Q^2)$ for a fixed $Q$ lead to weaker bounds. Making the very conservative assumption that the higher perturbative terms increase by a factor of 2 the value calculated from the sum in (\ref{eq:OPE}) truncated after four terms, we obtained the slightly larger interval $c_s^\pi\in (31-36.7) \gev^{-4}$.  

The bounds become weaker also if  the value of the spacelike energy $Q$ is increased (this  was already illustrated in Fig. \ref{fig:4}). Therefore, in order to obtain bounds of interest one should take the lowest $Q$ for which the perturbative expansion is considered to be reliable. The range (\ref{eq:c}) was obtained by assuming that this value is $Q=2 \gev$, which, as discussed in Sec. \ref{sec:input},   is reasonable  for light-quark correlators  evaluated on the spacelike axis. Assuming this value to be $Q= 4 \gev$ gives the larger interval $c_s^\pi\in (29-36) \gev^{-4}$, while the choice $Q=10 \gev$  leads to the even  larger range $c_s^\pi\in (17-46) \gev^{-4}$. In these calculations we used both scales $\mu=2 \gev$ and $\mu=Q$, which lead to very close results, and took the most conservative bounds.

Our analysis indicates that the model-independent bounds can accommodate the low values predicted by the older dispersive calculations only if  the corrections to the normalization of the form factor at $t=0$  are much larger than quoted in (\ref{eq:FH}), or the perturbative QCD regime for the scalar correlator is assumed to start at quite large energies.  Whether these strong assumptions are necessary or not remains to be established.

The results derived in this paper can be improved in principle by including in the unitarity sum for the spectral function $\mbox{\rm Im} \Psi(t)$, besides the $\pi\pi$ states, also the contribution of the $K\bar K$ states, nonzero for $t\ge 4 m_K^2$, which can be expressed in terms of the kaon scalar form factor $\Gamma_K(t)$. One can thus include information about $\Gamma_K(0)$ available in $\chi$PT, however, as discussed in \cite{DGL}, the knowledge of this quantity is not very precise. In addition, one has to evaluate also the contribution of the unphysical cut of $\Gamma_K(t)$ along $(4 m_\pi^2,\, 4 m_K^2)$,  which requires some model-dependent assumptions. Therefore, we shall not pursue this line here\footnote{The inclusion of higher states in the Meiman-Okubo formalism was used in  
\cite{Ananthanarayan:2014pta} for constraining the $\omega\pi$ form factor.}.

We  finally note that the uncertainties of the input quantities can be accounted for in a more realistic way by merging the present formalism with Monte Carlo simulations,  as done in the recent analysis  \cite{PRL:2017} of the pion charge radius.


\section*{Acknowledgments} 
I thank B. Ananthanarayan, Heiri Leutwyler, and Laurent Lellouch for  useful discussions and suggestions on the manuscript. This work was supported by the Romanian Ministry of Research and Innovation, Contract No. PN 18090101/2018.

\newpage

\end{document}